\documentstyle[multicol,amsfonts,aps,epsf]{revtex}
\newcommand{\beq}{\begin{equation}}
\newcommand{\eeq}{\end{equation}}
\newcommand{\beqa}{\begin{eqnarray}}
\newcommand{\eeqa}{\end{eqnarray}}

\begin{document}

\title{Interface Fluctuations under Shear}

\author{
Alan J. Bray,
Andrea Cavagna,
and
Rui D. M. Travasso
}

\address{
Department of Physics and Astronomy, University of Manchester, 
Manchester M13 9PL, United Kingdom}

\date{\today}

\maketitle

\begin{abstract}
Coarsening  systems under  uniform shear  display a  long  time regime
characterized by  the presence of  highly stretched and  thin domains.
The  question then  arises whether  thermal fluctuations  may actually
destroy this layered structure. To address this problem in the case of
non-conserved  dy\-na\-mics we  study  an anisotropic  version of  the
Burgers equation,  constructed to describe thermal  fluctuations of an
interface in the presence of a uniform  shear flow. As a  result, we find
that  stretched domains  are  only marginally  stable against  thermal
fluctuations in $d=2$, whereas they are stable in $d=3$.
\end{abstract}

\vskip 0.5 truecm

\begin{multicols}{2}

The  dynamics of  phase  separation in  systems  quenched below  their
critical temperature is substantially  modified when an external shear
is  applied. This  is true  both for  the case  of  conserved dynamics
(spinodal decomposition in  binary fluids), and non-conserved dynamics
(coarsening in Ising spin  systems).  In the simplest theoretical case
where  the sample  is  confined  between two  boundaries  moving at  a
constant relative velocity, intuition suggests that the growing domain
structure becomes  anisotropic, with  domains highly stretched  in the
direction  of the flow.   This fact  is now  well established  by many
experimental   \cite{experiment}    and   theoretical   investigations
\cite{theory,beppe-rapapa,orange,Ohta,beppe2}: it  has been shown that
the growth of the domains  is enhanced along the flow direction, while
it is less  clear whether the transverse growth  rate is unaffected by
the shear or depressed by it.  In either case, the long time effect of
the shear  is to form a  structure of very long  and (relatively) thin
domains.

A  natural question in  this context  is to  what extent  this layered
structure  is  stable  against  thermal fluctuations:  long  and  thin
domains may develop fluctuations,  transverse to their main axis, that
grow  large  enough to  break  them.  Similar  stretching-and-breaking
mechanisms have been proposed before \cite{Ohta,beppe2}, in particular
to argue  that the  anisotropic domain growth  may eventually  reach a
steady state  at late times  \cite{Ohta}.  However, the effect  of the
shear is  not only to  stretch the domains,  but also to  smooth their
surfaces,  and the  net  result of  the  two effects  is difficult  to
predict.  In  order to clarify this  problem a suitable  model for the
growth of a surface under shear must be analyzed.

Unfortunately, not many analytic  studies of domain growth under shear
exist.  In  particular, conserved dynamics  and spinodal decomposition
have  been analytically  considered mainly  in the  limit  of infinite
dimension of the order  parameter \cite{beppe-rapapa}, where no domain
interfaces are present. The situation is better in the simpler case of
non-conserved dynamics, where  zero-temperature coarsening under shear
has  recently been  studied analytically  \cite{orange}.  Our  task is
therefore to study  the effect of thermal fluctuations  in the case of
non-conserved dynamics  under shear. To  this end we will  introduce a
stochastic equation for a  scalar field $h(\vec x,t)$ representing the
height of  a fluctuating interface  above its flat ground  state. This
equation  will take  into account  the smoothing  effect of  the shear
flow, which  in fact  gives origin to  a non-linearity of  the Burgers
type.

Consider  an  interface separating  two  regions  with opposite  order
parameter.  When the temperature is  small but non-zero, there will be
some fluctuations, resulting in a non-flat profile parametrized by the
height $h(\vec x,t)$. In the  following, we will indicate with $d$ the
total  dimension of  the space,  and with  $d'$ the  dimension  of the
substrate spanned  by $\vec  x$, that is  $d=d'+1$. In the  absence of
shear,  the  free  energy cost  of  a  non-flat  profile is  given  by
$F=(\sigma/2)\, \int  d\vec x \, (\vec\nabla h)^2$,  where $\sigma$ is
the surface tension, and the corresponding Langevin equation coincides
with  the standard Edwards-Wilkinson  (EW) growth  equation \cite{EW},
\beq 
\partial_t h = \nu \nabla^2 h +\xi \ ,
\label{EW}
\eeq 
where $\nu$ is a diffusion coefficient (equal to $\sigma$ divided
by a  kinetic coefficient) and  $\xi(\vec x,t)$ is  a delta-correlated
noise,  
\beq   
\langle  \xi(\vec  x,t)\xi(\vec   x',t')\rangle  
=  D\,\delta(\vec x-\vec x')  \delta(t-t') \ .  
\eeq 
The  noise strength $D$
is  proportional to  the temperature.  As is  well known,  due  to its
linearity a simple scaling analysis  of the EW equation gives the exact
critical exponents. Under the rescaling $x\to bx, \ t\to b^{z_0}t$ and
$h\to  b^{\chi_0}  h$,  we  have  $\nu\to  b^{z_0-2}\nu  $  and  $D\to
b^{z_0-2\chi_0-d+1}  D$.  By  imposing  scale  invariance,  we  obtain
$z_0=2$  and $\chi_0=(3-d)/2$.  Even  though the  EW  equation is  not
suitable  for  describing  interface  fluctuations  when  a  shear  is
present, it is interesting to  see what the EW exponents would predict
for  the   stability  of  the   stretched  domains.   For   $d=2$  the
investigation of  \cite{orange} gives two length  scales, $L_{\|}(t) =
O(t)$ and $L_\bot  = O(1)$ (up to logarithmic  factors). Within the  
context of the EW equation the transverse fluctuations will grow as 
\beq
h \sim t^{\chi_0/z_0} F(t/L_{\|}^{z_0}),
\label{transverse}
\eeq
where $F$ is a scaling function with the limiting forms 
$F(0)={\rm const}$, $F(s) \sim s^{-\chi_0/z_0}$ for $s \to \infty$. 
In writing down (\ref{transverse}) we have assumed that $L_{\|}$ is a 
fixed length scale, while actually it is growing with time. The 
interpretation of (\ref{transverse}) is, however, simple. If 
$t \ll L_{\|}^{z_0}$, the domains are coarsening faster than the 
interfacial fluctuations, and we can effectively set $L_{\|}$ to 
infinity. Then $h \sim t^{\chi_0/z_0}$. On the other hand, if 
$t \gg L_{\|}^{z_0}$, the fluctuations are coarsening faster than 
the domains and eventually equilibrate on the scale $L_{\|}$. This 
corresponds to the large argument limit of the scaling variable in 
(\ref{transverse}), giving $h \sim L_{\|}^{\chi_0}$. Combining these 
two limits gives
\beq
h \sim {\rm min}\,\left(t^{\chi_0/z_0},  L_{\|}^{\chi_0}\right)\ .
\label{min}
\eeq
In $d=2$, this gives $h \sim t^{1/4}$, while $L_\bot = 0(1)$. 
Thus,  thermal fluctuations would eventually disrupt  the domain  
structure  and destroy  the  coarsening state. In $d=3$, on the  
other hand, it has  been found  that \cite{orange} $L_\|(t) =
O(t^{3/2})$ and  $L_\perp(t)=O(t^{1/2})$. In this case, $\chi_0=0$ 
(logs), so  conventional  EW  thermal  roughening would  not  destroy  the
domains, since  $h \ll  L_\bot$. 
As we  shall see, the  inclusion  of shear  in  equation  (\ref{EW})  
will modify  these results. In particular, the instability of the domains 
in $d=2$ will be reduced, while stability in $d=3$ will be confirmed.

When a shear flow is present the EW equation must be modified. We will
consider a standard  shear velocity profile $\vec u$,  with flow along
the $x$ direction and shear gradient perpendicular 
to the surface. Let us label this last direction by $z$, such that, 
\beq \vec u=\gamma\, z\, \vec e_x \ ,
\label{flowba}
\eeq 
where $\gamma$ is the shear  rate. In this way we break 
the symmetry between the $x$ direction and the remaining $(d'-1)$ 
directions $\vec x_\perp$. Note that the growth is orthogonal to 
the shear flow: this is what happens to the domain walls 
in the  long-time limit of a coarsening process under shear. The 
field $h(\vec x,t)$ is now dragged in the $x$ direction by an  amount 
proportional to $h$ itself, that is 
$\partial_t \to\partial_t + \gamma h\partial_x $. The correct equation
for $h(\vec x,t)$ therefore becomes,
\beq \partial_t h  + \gamma\, h\,\partial_x  h  
=  \nu_x  \partial_{xx} h +    
   \nu_\perp  \nabla_\perp^2  h + \xi  \ ,
\label{lei}
\eeq
where  we have  introduced separate  diffusion constants,  $\nu_x$ and
$\nu_\perp$.
For $d'=1$ ($d=2$) equation (\ref{lei}) is  nothing other 
than the  Burgers equation \cite{burgers}. This can be  mapped 
onto the  KPZ equation   \cite{kpz} with   space-correlated   noise,    
via   the transformation   $h=\partial  _x\hat   h$,  i.e.\  
$h\,\partial  h\to \frac{1}{2}\partial(\partial\hat h)^2$, yielding  
the KPZ equation for $\hat{h}$.  Such  a case  has been studied  in 
\cite{medina}.  In generic dimension equation (\ref{lei}) is anisotropic
and was first introduced in \cite{kardar}, in the context of a 
model of sandpiles, and further studied in \cite{janssen}.
A standard method for the analysis of stochastic non-linear equations
is the dynamic renormalization group (RG) approach, first used in 
this context in \cite{fns}. Here, we will briefly review those RG 
results for equation (\ref{lei}) which are most relevant for our purpose. 
We start the analysis of equation (\ref{lei}) 
by finding the bare  scaling dimensions  of  
the parameters.   Under the  anisotropic rescaling, 
\beq 
x\to b x \ \ ,\ \vec x_\perp \to b^\zeta \vec x_\perp \ \ ,\ t
\to b^z t \ \ , \ h \to b^\chi h \ ,
\label{scala}
\eeq 
we obtain, 
\beqa
\nu_x & \to & b^{z-2}\ \nu_x\  , \nu_\perp \to
b^{z-2\zeta} \ \nu_\perp\ , \gamma \to b^{\chi+z-1} \ \gamma
\ ,\nonumber \\ 
D & \to & b^{z-2\chi -1 -(d-2)\zeta} \ D \ .
\label{riscala}
\eeqa 
Inserting the EW exponents $\chi_0$ and $z_0$ found above for  
$\gamma=0$  into the scaling equation for $\gamma$ gives: 
$\gamma \to b^{(5-d)/2}\,\gamma$,  showing that $d_c = 5$ is 
the critical 
dimension below which the nonlinearity becomes relevant. Thus, for 
$d<5$ the EW exponents are no longer correct and an RG
approach becomes necessary.

Burgers  equation is  known  to  be  invariant under  a  Galilean
transformation.  In  the  present context,  we have  an
equivalent  symmetry, namely  the invariance  of  equation (\ref{lei})
under a coordinate transformation which preserves the original form of
the shear flow. Indeed, from (\ref{flowba}) we have $x\sim\gamma h t$,
and therefore if we vertically  shift the interface, $h\to h+h_0$, the
equation is  invariant provided that we make  the transformation $x\to
x+\gamma h_0  t$. Basically, this is just  translational invariance in
the  $z$ direction.  This  symmetry is  important, because  it implies
that $\gamma$ cannot be perturbatively  corrected in a RG analysis 
and therefore we must set to zero its bare scaling dimension. In this way
a relation between dynamic and growth exponents is obtained, $\chi+z=1$.
The remarkable feature of equation (\ref{lei}) is that there are two  
further parameters whose bare scaling dimensions are not perturbatively 
changed, namely $\nu_\perp$ and $D$. 
This happens because  the $\gamma$-vertex in equation (\ref{lei}) 
carries in Fourier space a factor $k_x$ that  can
generate  neither  terms  $O(k_\perp^2)$ in the self- energy, 
contributing to a renormalization of  $\nu_\perp$, nor terms  $O(1)$ 
in the renormalized  noise, contributing  to  the  renormalization  of
$D$ \cite{kardar}. 
We can therefore set to zero the bare scaling dimensions 
of $\nu_\perp$ and $D$, obtaining in this way  two  extra  
relations among  the  fixed  point
exponents: $z=2\zeta$ and $z=2\chi+1+(d-2)\zeta$. 
Therefore, for $d\leq 5$, we have \cite{kardar}, 
\beq 
z = \frac{6}{8-d} \quad ,\quad 
\chi = \frac{2-d}{8-d} \quad , \quad 
\zeta = \frac{3}{8-d} \ .
\label{result3d}
\eeq 
From (\ref{scala}) and (\ref{result3d}) we can infer the scaling form for 
the correlation function in Fourier space, defined by 
$\langle h(\vec{k},\omega) h(\vec{k}',\omega')\rangle
=C(\vec{k},\omega)\delta(\vec{k}+\vec{k}')
\delta(\omega+\omega')$. We have,
\beq
C(\vec{k},\omega) = \frac{1}{k_x^{2z}}\,f\left( \frac{\omega}{k_x^z},
\frac{|\vec{k}_\bot|}{k_x^\zeta}\right) \ ,
\label{scalingfn}
\eeq
where $f$ is a scaling function.

It is useful to rescale coordinates and field in order to
identify the effective coupling constant of equation 
(\ref{lei}). The change of variables,
\beqa
h &\to & \left(D/\nu_x^{2-d/2}\nu_\perp^{d/2-1}\right)^{1/2}h,
\ \ x_\perp \to \left(\nu_\perp/\nu_x\right)^{1/2}x_\perp, \nonumber \\
t &\to & (1/\nu_x)\,t,
\ \ \xi \to \left(D \nu_x^{d/2}/\nu_\perp^{d/2-1}\right)^{1/2}\xi\ ,
\label{eff}
\eeqa
amounts to setting $\nu_x=\nu_\perp=D=1$ and replacing the vertex 
$\gamma$ by the effective vertex
\beq
\hat\gamma=\frac{\gamma \ D^{1/2}}{\nu_x^{2-d/4} \nu_\perp^{d/4-1/2}} \ .
\eeq
As expected, for $d=d_c=5$ this quantity is dimensionless.
Using standard RG  methods (see, for example, \cite{fns,kardar,medina})
it is possible to obtain the one-loop RG 
flow equation for the effective coupling constant
$U=c \,\hat\gamma^2$, where  
$c=\Gamma\left(3-d/2\right)/[8(4\pi)^{d/2-1}]$. The flow equation 
reads:
\beq
\frac{d U}{d l} =  
(5-d)\ U - \frac{1}{2}(8-d)^2\ U^2 \ .
\label{flow}
\eeq
The trivial fixed point, $U=0$, becomes stable 
above the critical dimension $d_c=5$, giving the EW 
exponents $z_0,\ \chi_0$, corresponding to $\zeta=\zeta_0=1$. 
No other physical $(U>0)$ fixed point exists in this phase. The 
effect of the shear on thermal 
roughening is therefore negligible above dimension $d_c=5$ and 
domain wall fluctuations are isotropic (in the $d'$-dimensional 
subspace parallel to the mean orientation of the wall).
On the other hand, in dimension $d<5$ there is an attractive fixed 
point at $U=2\,\epsilon/9 + O(\epsilon^2)$, with $\epsilon=d_c-d$. 
The effective coupling constant is of order $\epsilon$, such
that close to the critical dimension the RG expansion is  
under control. This fixed point controls the large-scale behavior of 
equation (\ref{lei}) for $d<5$ and is associated with the non-trivial 
exponents (\ref{result3d}).

The scaling exponent $\chi$ of the interface height is zero for
$d=2$, while it is negative, $\chi=-1/5$, for $d=3$. From equation 
(\ref{min}), with $\chi_0$ and $z_0$ replaced by the new exponents 
$\chi$ and $z$, we see that in the  two-dimensional  case  the  highly  
elongated  domains  are  only marginally   stable   against  thermal 
fluctuations, since $L_\parallel^\chi=O(1)$, $t^{\chi/z}=O(1)$, and 
$L_\perp=O(1)$ (up to logarithms in each case).  A thermally-induced 
stretching-and-breaking mechanism cannot, therefore, be excluded in 
this case. On the other hand, for $d=3$ a negative value of $\chi$ 
means that thermal fluctuations of the interfaces saturate at late 
times. Thus, in  the  three-dimensional case thermal roughening is 
depressed by the shear and the  flow-induced layered  structure  
of  the domains  is stable  against  thermal   fluctuations.

In this work we have shown that interface fluctuations under shear,
in a system with non-conserved order parameter, can be described 
by a stochastic differential equation, with an anisotropic Burgers 
non-linearity, and that the critical exponents are known exactly.
In this way it was possible to assess the stability of domains
in the late time regime of a system subject to a uniform shear flow.

AB thanks R.K.P. Zia for a useful discussion. 
This work was supported by EPSRC  grant GR/L97698
(AC and AJB), and by Funda\c c\~ao para a Ci\^encia e a Tecnologia
grant BD/21760/99 (RDMT).

\end{multicols}

\end{document}